\documentclass[12pt,twoside]{article}
\usepackage{a4}\usepackage{epsf}
\usepackage{epsf}\usepackage{latexsym}
\def \lsim{\,{\scriptscriptstyle{\stackrel{<}{\sim}}}\,}

\newcommand{\be}{\begin{equation}}
\newcommand{\ee}{\end{equation}}
\newcommand{\bea}{\begin{eqnarray}}
\newcommand{\eea}{\end{eqnarray}}
\newcommand{\beq}{\begin{eqnarray}}
\newcommand{\eeq}{\end{eqnarray}}
\newcommand{\beao}{\begin{eqnarray*}}
\newcommand{\eeao}{\end{eqnarray*}}
\newcommand{\nn}{\nonumber}
\newcommand{\pa}{\partial}
\newcommand{\e}{{\rm e}}

\renewcommand{\d}{{\rm d}}
\newcommand{\Tr}{{\rm Tr~}}

\newcommand{\Ref}[1]{(\ref{#1})}
\newcommand{\al}{{\alpha}}
\newcommand{\ep}{{\epsilon}}
\newcommand{\me}{m_{\rm e}}
\newcommand{\bc}{boundary conditions }

\newcommand{\gse}{ground state energy }
\newcommand{\gsep}{ground state energy. }
\newcommand{\gsek}{ground state energy, }
\newcommand{\GSE}{Ground state energy }
\newcommand{\hke}{heat kernel expansion }
\newcommand{\hkep}{heat kernel expansion. }
\newcommand{\hkk}{heat kernel coefficient }
\newcommand{\hkks}{heat kernel coefficients }

\renewcommand{\P}{{\cal P} }

\newcommand{\E}{{\cal E}}

\newcommand{\dpe}{\Delta{\cal P} (r_1)}
\newcommand{\dpz}{\Delta{\cal P} (r_2)}
\newcommand{\dpd}{\Delta{\cal P} (r_3)}
\newcommand{\dpv}{\Delta{\cal P} (r_4)}
\newcommand{\fhet}{\Phi_{H^1} (r_1)}
\newcommand{\fhzt}{\Phi_{H^1} (r_2)}
\newcommand{\fhdt}{\Phi_{H^1} (r_3)}

\newcommand{\fhe}{\Phi_{H^1}^T (r_1)}
\newcommand{\fhz}{\Phi_{H^1}^T (r_2)}
\newcommand{\fhd}{\Phi_{H^1}^T (r_3)}
\newcommand{\fhv}{\Phi_{H^1}^T (r_4)}
\newcommand{\fje}{\Phi_{J} (r_1)}
\newcommand{\fjz}{\Phi_{J} (r_2)}
\newcommand{\fjd}{\Phi_{J} (r_3)}
\newcommand{\fjv}{\Phi_{J} (r_4)}

\newcommand{\fjzt}{\Phi_{J}^T (r_2)}
\newcommand{\fjdt}{\Phi_{J}^T (r_3)}
\newcommand{\fjvt}{\Phi_{J}^T (r_4)}
\newcommand{\intre}{\int_0^{\infty}dr_1\,r_1\,}
\newcommand{\intrz}{\int_0^{r_1}dr_2\,r_2\,}
\newcommand{\intrd}{\int_0^{r_2}dr_3\,r_3\,}
\newcommand{\intrv}{\int_0^{r_3}dr_4\,r_4\,}
\newcommand{\pre}{\left(\frac{\pi}{2i}\right)}
\begin{document}
\title{The ground state energy of a spinor field in the background 
of a finite radius flux  tube} 
\author{{\sc 
M. Bordag\thanks{e-mail: Michael.Bordag@itp.uni-leipzig.de} and
K. Kirsten}\thanks{e-mail:  Klaus..Kirsten@itp.uni-leipzig.de}\\ \\
  Universit\"at Leipzig, Fakult\"at f\"ur Physik und Geowissenschaften \\
  Institut f\"ur Theoretische Physik\\
  Augustusplatz 10/11, 04109 Leipzig, Germany}
\maketitle
\begin{abstract}
We develop a formalism for the calculation of the ground state energy
of a spinor field in the background of a cylindrically symmetric
magnetic field.  The energy is expressed in terms of the Jost function
of the associated scattering problem. Uniform asymptotic expansions
needed are obtained from the Lippmann-Schwinger equation. The general
results derived are applied to the background of a finite radius flux
tube with a homogeneous magnetic field inside and the ground state energy is
calculated numerically as a function of the radius and the flux. It
turns out to be negative, remaining smaller by a factor of $\alpha$
than the classical energy of the background except for very small
values of the radius which are outside the range of applicability of QED.
\end{abstract}
\thispagestyle{empty}\setcounter{page}{0}
\newpage\setcounter{page}{1}
\section{Introduction}\label{Sec1}
The \gse of the spinor field in the background of a magnetic field had been
investigated since the early days of Quantum Electrodynamics (QED).
So, for instance, the effective potential in a strong magnetic field is well
known.  For a weak field it takes increasing positive values, for stronger
fields it turns down and for $B\to\infty$ it becomes
\be\label{ash}
V_{\rm eff} \sim -{(eB)^{2}\over 12\pi^{2}} \ln {\sqrt{2eB}\over \me}\, ,
\ee
where $\me$ is the electron mass. For details see e.g. \cite{dtz79}.  

Actual interest in this topic results, for example, from the symmetry
restoration due to a magnetic field in the electroweak theory or from the
influence of that field on the character of the phase transition
\cite{elmfors}. While most work had been done in homogeneous magnetic fields,
mainly because in that case explicit formulas exist, the extension to
inhomogeneous fields is of interest.

Some work is this direction has already been done. For example for a flux
where the magnetic field is concentrated on the surface of the tube, the
fermion determinant was calculated in \cite{fry}. However, in that case the
classical energy is infinite.
In \cite{dunne}, in (2+1) dimensions a magnetic field homogeneous in the
$y$-direction and with a special shape in the $x$-direction allowing for
explicit formulas had been considered.  The result for the \gse per unit area
of the planes was expressed in quite elementary functions, which allowed the
discussion of the total energy within the family of fields considered. It was
shown, that the system is driven towards a uniform magnetic field.  An
extension of these formulas to the (3+1) dimensional case was given in
\cite{dunne98}.

Furthermore, there exists a number of investigations of the density per unit
volume $\ep(r)$ of the \gsek where $r$ is the radial coordinate in the
perpendicular plane, in the background of an infinitely thin magnetic flux
tube. Partly, this is motivated by the close relation to the Aharonov-Bohm
effect. The first investigation of this kind was done in \cite{serebryany} for
the \gse density of a scalar field, later reconsidered and generalised to the
spinor case in \cite{g90}, see also \cite{sitenko}. Extensively investigated
is the (2+1)-dimensional case, see \cite{lcg93} and papers cited
therein. Also, there are similar investigations in the background of an
infinitely thin cosmic string \cite{bt95}.
The calculations in the background of an infinitely thin magnetic flux tube
have the drawback that the energy density per unit volume cannot be integrated
to get the energy density $\E$ per unit length due to the singular behaviour
near the string, $\ep(r)\sim r^{-4}$, which follows already from dimensional
reasons.

In addition one might consider the combined effect of boundaries and
background fields. This has been started in \cite{mariel}, where imposing
spectral boundary conditions the spinor field was considered in a finite
region of space in the background of an Aharonov-Bohm flux string.

In the present paper we continue the consideration of inhomogeneous magnetic
background fields and calculate the \gse of the spinor field in the background
of a straight magnetic flux tube of {\bf finite} radius $R$, more exactly, the
energy density per unit length. The reason to consider a flux tube of finite
radius $R$ is that the associated classical energy is finite and the
dependence of the total energy when $R$ varies while the flux is fixed can be
analysed.  The interesting question in this context is if some radius $R_m$
exists where the complete energy, i.e., the sum of classical enegy of the
magnetic field and the \gse of the spinor field, is minimized and the magnetic
string gets stable. In doing this analysis we use and generalise the formalism
developed in \cite{b95} and \cite{bk96} for a smooth scalar background field.
In section 2 we start discussing in detail the renormalization of the ground
state energy. We will normalize the energy in such a way that it vanishes for
the electron mass $m_e \to \infty$, the resulting massless limit will also be
considered.  The needed counterterms and the subtraction employed is then
elegantly described using the heat-kernel language. After having explained in
detail the renormalization procedure we express the ground state energy in
terms of the Jost function of the associated scattering problem. The procedure
developed in \cite{b95,bk96} consists of adding and subtracting the uniform
asymptotic expansion of the Jost function. This is done in section 4 using a
perturbative expansion of the Lippmann-Schwinger equation. Various details of
this calculation are relegated to the Appendices A and B. In the remaining
part of the paper, section 5 and 6, an analytical as well as numerical
description of the ground state energy for spinors in the presence of a finite
radius flux tube is performed.  Some details can be found in Appendix C.  In
the conclusion we summarize the main results of the paper.

\section{Basic formulas and the renormalization}\label{Sec2}
The considered background is a straight magnetic flux tube of finite radius
$R$, i.e., the magnetic field
\be\label{mf}
\vec{B}(\vec{x})=\frac\phi{2\pi} \ h(r) \ \vec{e}_{z}\, ,
\ee
where $h(r)$ is a profile function with compact support in the radial variable
$r=\sqrt{x^{2}+y^{2}}$ in the plane perpendicular to the tube. By the
normalization $\int_{0}^{\infty}\d r r h(r)=1$, $\phi$ has the meaning of the
flux inside the tube. The corresponding vector potential can be chosen to be
\be\label{ma}
\vec{A}(\vec{x})={\phi\over 2\pi} \ {a(r)\over r} \ \vec{e}_{\varphi}.
\ee
The profile functions $h(r)$ and $a(r)$ are connected by the relation
$h(r)=a'(r)/r$. Below, we will use the special case of a homogenoeus magnetic
field inside the tube, where these functions read
\be\label{mah}
h(r)={2\over R^{2}}\Theta(R-r), \qquad a(r)={r^{2}\over
  R^{2}}\Theta(R-r)+\Theta(r-R). 
\ee
In this case the solutions of the field equations can be expressed in terms of
Bessel and hypergeometric functions. 

Another choice could be a magnetic field concentrated on the surface of the
cylinder, $h(r)=\delta(r-R)/R$, where the solutions can solely be expressed in
terms of Bessel functions. But in that case the classical energy of the
background is infinite. 

The classical energy of the background (per unit length of the string) is
\be\label{eclass}
\E^{\rm class}\equiv\frac12\int\d\vec{x} \ \vec{B}^{2}
={\phi^{2}\over 4\pi}\int\limits_{0}^{\infty}\d r \ r \ h(r)^{2}.
\ee
The \gse of the spinor field in that background is given by
\be\label{gze}
\E=-{\mu^{2s}\over 2} \ \sum\limits_{(n,\ep)} \ {e_{(n,\ep)}}^{1-2s}.
\ee
Here, the minus sign in front of the rhs. accounts for the spinor 
obeying anticommutation relations, 
$s$ ($s>2$ and $s\to 0$ in the end) is the regularisation parameter
in the zetafunctional regularisation which we use here and $\mu$ is the 
arbitrary dimensional parameter entering this regularisation.  In fact, due to
the translational invariance along the axis of the flux tube, we have to
consider the energy density per unit length. This will be taken into account
below. 

The
$e_{(n,\ep)}$ are the eigenvalues of the Hamiltonian
\be\label{p} {\cal H}=-i\gamma^{0}\gamma^{j}\left(\pa/\pa
x^{j}-ieA_{j}(x)\right)+\gamma^{0}\me
\ee
which follows from the Dirac equation.  Here, $\ep=\pm 1$ is the sign of the
one particle energies $e_{(n,\ep)}$ for the particle respectively
the antiparticle
which themselves are chosen to be positive. 
All other quantum numbers are
included into $(n)$.
Furthermore, the index $j$ denotes
spatial indices only and summation over it is included.

For the renormalisation we follow the standard procedure using the \hkep 
The \gse can be expressed by the keat kernel $K(t)$ of ${\cal H}^2$, 
\[
\E=-{\mu^{2s}\over 2}\int\limits_{0}^{\infty}{\d
  t~t^{s-3/2}\over\Gamma(s-1/2)}~K(t)
\]
with the asymptotic expansion for $t\to 0$
\[
K(t)
\sim {\e^{-tm^{2}}\over (4\pi t)^{d/2}}\sum\limits_{n\ge 0}a_{n}t^{n},
\]
where $d$ is the dimension of the manifold, $d=3$ in our case.

The heat-kernel coefficients $a_n$ for the operator 
under consideration
are well known. The relevant operator ${\cal H}^2$ reads explicitly,
\beq
{\cal H}^2 = -\nabla^j \nabla_j+
\frac12 \sigma^{ij} F_{ij} +m^2 , \nn
\eeq
with $\sigma^{\mu\nu}=\frac{i}{2} [\gamma^{\mu}\gamma^{\nu}]$ and the leading
coefficients can be found for example in \cite{peter}.

For $n=0$ we note that the coefficient is independent on the
background field and corrresponds to the contribution of the empty
Minkowski space. We drop this contribution without further comment.
The coefficient $a_{1}$ is zero and for $a_{2}$ the general formula
reads
\be\label{a2allg}
a_{2}=\Tr \int\d \vec{x} \left(-\frac{1}{12} F_{ij}^{2}+\frac18
(\sigma^{ij}F_{ij})^{2}\right).
\ee
Here, the trace is over the spinors. The integration along the axis of
the flux tube gives the corresponding volume by which we have to
divide.  So the following formulas have to be understood always as
densities with respect to this axis.

The trace in (\ref{a2allg}) can be carried out
and by means of \Ref{ma} we arrive at
\be\label{a2a} a_{2}={8\pi\over 3}\delta^{2}\int\limits_{0}^{\infty}
{\d r ~r}~ h(r)^{2},  \ee
where the notation
\be\label{delta}
\delta=\left({e\phi\over  2\pi}\right)
\ee
is introduced. Here, $e$ is the electron charge, and we can rewrite this
relation as $\delta={\sqrt{\al}}\phi/(2\pi)$, where $\al$ is the fine
structure constant.

Using the \hke  it can be shown that the divergent part of the \gse  
results from the contribution of the \hkks $a_{n}$ with $n\le 2$. We define
\be\label{ediv}
\E^{\rm div}={a_{2}\over32\pi^{2}} \ \left(\frac{1}{s} -2 +\ln
  {4\mu^{2}\over\me^{2}}\right) \, .
\ee
Then the renormalised \gse is given by
\be\label{eren}
\E^{\rm ren}=\E-\E^{\rm div}.
\ee
Here the limit $s\to 0$ can be performed because the pole part is subtracted.
In general, the definition of $\E^{\rm div}$ is not unique.  By the definition
\Ref{ediv}, the normalisation condition
\be\label{normcond}
\E^{\rm ren}\to 0 ~~~~~~~~~~~~~~ \mbox{for} ~~~~~~~~\me \to \infty
\ee
is assured.  This normalisation condition is natural as it implies that a very
massive field should not show quantum fluctuations. On the other hand, it
fixes the arbitrariness which came in with the parameter $\mu$ in \Ref{gze}.

Along with the subtraction of $\E^{\rm div}$ from $\E$ it must be
added to $\E^{\rm class}$. This is equivalent to a renormalisation of the
flux according to
\be\label{fren} 
\phi^{2} \to \phi^{2}+ {\left({e\phi}\right)^{2}\over 12\pi^2}
\left(\frac{1}{s} -2 +\ln {4\mu^{2}\over\me^{2}}\right) \, .  
\ee

{}From this renormalisation procedure it is possible to determine the leading
asymptotic behaviour of the renormalised \gse when the radius $R$ of the flux
tube tends to zero. In fact, this could have been done using the arguments
given in \cite{bvw} where the general scaling behaviour of the Casimir energy
was investigated. The point is simply that the regularized \gse \Ref{gze} has
a series expansion with respect to powers of the mass $\me$. Note that this is
not affected by the zero modes, which are present here (see \cite{ac}),
because we consider the \gse and not the determinant of the operator $\P$.
Then, by means of \Ref{eren} and \Ref{ediv} we subtract a contribution
containing $\ln \me$. Therefor, the renormalized \gse $\E^{\rm ren}$ becomes
for $\me\to 0$ proportional to $\ln \me$. Now, for dimensional reasons it can
be written as
\[
\E^{\rm ren}={f(R\me)\over R^{2}}
\]
where $f$ is some function of the dimensionless combination $R\me$.
Consequently, for $R\to 0$ the behavior must be
\[
\E^{\rm ren}\sim {-a_{2}\over 16\pi^{2}}\ln{1\over R\me}
\]
and, for instance, with the background \Ref{mah}
\be\label{asmod}
\E^{\rm ren}\sim {-1\over 3\pi}{\delta^{2}\over R^{2}} \ \ln{1\over
  R\me}\ .
\ee
This behavior will be confirmed below in the course of the explicit
calculations.  In fact, this behavior follows already from the \hke and has
the same origin as the asymptotics \Ref{ash}. For the same reason it can
obviously be improved using the renormalisation group. Because the theory is
infrared free and the limit $R\to 0$ corresponds to high momenta it is
impossible to make $R$ too small.  

When comparing \Ref{asmod} with the classical energy of this background,
\be\label{eclass1}
\E^{\rm class}={\phi^{2}\over 2\pi R^{2}}\ ,
\ee
one could think that the complete energy can be made negative for
sufficiently small $R$. However, this would require $ R\me < \exp
(-6\pi^{2}/\al)$, which is far outside the range of applicability of
QED and also ruled out by the renormalization group argument.

Note that in the massless case the renormalization scheme must be
modified. As discussed in \cite{bkv98}, for $a_2\ne 0$ which is the
case here generically ($a_2=0$ means a vanishing background,
cf. \Ref{a2a}) and which represents the conformal anomaly, there is no
transition for the renormalised \gse $\E^{\rm ren}$ from $\me\ne 0$ to
$\me =0$. Usually, in the massless case   in the definition of $\E^{\rm
div}$ like \Ref{ediv} instead of the mass, another parameter, the
renormalization scale $\Lambda$, is used. Therefor the renormalised
\gse contains a nonuniqueness proportional to the \hkk $a_2$, i.e.,
proportional to the classical energy (see formulas \Ref{eclass} and
\Ref{a2a}).

\section{\GSE expressed in terms of the Jost function}\label{Sec3}
We express the regularised \gse \Ref{gze} in terms of the Jost function of the
scattering problem associated with the operator ${\cal H}$,
Eq. \Ref{p}. Because the
background is translationally invariant along the third axis, by means of
\[
\Psi_{(n,\ep)}(\vec{x})=\e^{ip_{3}x^{3}}\left({\Phi\atop\psi}\right) ,
\]
we rewrite the corresponding Dirac equation using the chiral representation of
the gamma matrices in the form
\be\label{eq1}
\left(\begin{array}{rcl}
p_{0}+\hat{L}-\me \sigma_{3} & & p_{3}\sigma_{3}\\
p_{3}\sigma_{3}& &p_{0}+\hat{L}+\me \sigma_{3} \end{array}\right)
\left({\Phi\atop\psi}\right)=0 ,
\ee
with $\hat{L}=i\sum_{i=1}^{2}\sigma_{i}(\pa/\pa x^{i}+ieA_{i})$. It is
sufficient to consider $p_{3}=0$ and one of the two decoupled equations. Then
by means of
\[
\Phi=\left(\begin{array}{rl}
ig_{1}(r) & \e^{-i(m+1)\varphi}\\ g_{2}(r) & \e^{-im\varphi}\end{array}\right)
\]
$(m=-\infty,\infty)$ we arrive at the equation
\be\label{eq2}
\left(\begin{array}{rcl}
p_{0}-\me  & & {\pa\over \pa r}-{m-\delta a(r)\over r}\\
- {\pa\over \pa r}-{m+1-\delta a(r)\over r}& &p_{0}+\me
\end{array}\right)\Phi(r)=0\ ,
\ee
where we introduced the notation
\be\label{lsg}
\Phi(r)=
\left({g_{1}(r)\atop g_{2}(r)}\right)
\ee
for the solution $\Phi (r)$.

The solutions in the exterior space, $r>R$, are
\be\label{fsol}
\Phi_{J}^{0}(r)
=\left\{
\begin{array}{rcl}
\left({\begin{array}{rl}\sqrt{p_{0}+\me}&J_{m-\delta+1}(kr)\\
      \sqrt{p_{0}-\me}&J_{m-\delta}(kr)\end{array}}\right) 
& \mbox{for} & m+1-\delta>0    \\[14pt]
\left({\begin{array}{rl}\sqrt{p_{0}+\me}&J_{\delta-m-1}(kr)\\
      -\sqrt{p_{0}-\me}&J_{\delta-m}(kr)\end{array}}\right) 
& \mbox{for} & m-\delta<0 \ .
\end{array} \right.
\ee
The Jost solution of Eq. \Ref{eq1} is the solution which behaves for
$r\to 0$
as the free solution \Ref{fsol}.  Its asymptotics for $r\to\infty$ can be
written as
\be\label{asj} \Phi(r)\sim
\frac12\left(f_{m}(k)\Phi^{0}_{H^{(2)}}(r)
  +\bar{f}_{m}(k)\Phi^{0}_{H^{(1)}}(r)\right), \ee
where $\Phi_{H^{(1,2)}}$ are the solutions \Ref{fsol} with the Hankel
functions instead of the Bessel function. The coefficient $f_{m}(k)$ is
the
Jost function and $\bar{f}_{m} (k)$ its complex conjugate.
 
The  \gse can be expressed  in terms of the Jost
function much in the same way as in the scalar case \cite{bk96}. However, due
to the translational invariance in the direction parallel to the flux tube, we
have the energy density
\be\label{gzestring}
\E=-{\mu^{2s}\over 2} \ \int\limits_{-\infty}^{\infty}{\d k_{3}\over 2\pi}
\sum\limits_{(n,\ep)} \ \left(k_{3}^{2}+e^2_{(n,\ep)}\right)^{1/2-s},
\ee
instead of the general formula \Ref{gze}.  After carrying out the integration
over $k_{3}$ we arrive at
\[
\E=-{\mu^{2s}\over 2}
\frac 1 {2\sqrt{\pi}} \frac{\Gamma (s-1)}{\Gamma (s-1/2)}
\sum\limits_{(n)} \left(e^2_{(n)}\right)^{1-s},
\]
where $(n)$ denotes the remaining
quantum numbers in the plane perpendicular to the axis of the tube. 
Now the sum over these $(n)$ can be expressed through the Jost function
and we get to the relevant order in $s$,
\be\label{eregj0}
\E=C_{s} \sum\limits_{m=-\infty}^{\infty}\int\limits_{\me}^{\infty}\d k \
(k^{2}-\me^{2})^{1-s} \ {\pa\over\pa k} \ \ln f_{m}(ik)
\ee
with $C_{s}=(1+s(-1+2 \ln (2\mu)))/(2\pi )$.  This representation can
be obtained in much the same way as done in the scalar case in
\cite{bk96}. One has to take into account the known analytical
properties of the Jost function (which differ from that in the scalar
case). One has to use as an intermediate step a finite quantization
volume with appropriate \bc (bag conditions work well). Then, in the
course of tending this volume to infinity, the translational invariant
contribution from the Minkowski space must be dropped and the remaining
finite part, after a deformation of the integration contour, just
delivers \Ref{eregj0}.

We remark that in the considered problem there are zero modes \cite{ac} (at
$k=0$).  In the sum in Eq. \Ref{gzestring} they have to be taken into account.
Just in the same way as shown in detail in the scalar case in \cite{bk96} for
the bound states they do not show up explicitely in representation Eq.
\Ref{eregj0}.

Here we have taken into accout that both signs of the one particle energies as
well as both signs of the spin projection give equal contributions to the \gse
thus resulting in a factor of 4 which is included into $C_{s}$. 
This expression will be used in the calculations below.

The renormalisation of the \gse is defined by Eq. \Ref{eren}. The remaining
task is to perform the analytical 
continuation as $s\to 0$. However, 
this is not immediately possible
using
representation \Ref{eregj0} for $\E$. 
To continue we use the uniform asymptotic expansion
$\ln f_{m}^{\rm as}(ik)$ of the logarithm of the Jost function, $\ln
f_{m}(ik)$, defined in such a way that the difference
\be\label{diff}
\ln f_{m}(ik)-\ln f_{m}^{\rm as}(ik) = O\left(\frac1{m^{4}}\right)
\ee
is of the order $m^{-4}$ in the limit $m\to\infty$, $k\to\infty$ for
$\frac{m}{k}$ fixed.
Then we split the renormalised \gse by adding and subtracting $\ln f_{m}^{\rm
  as}(ik)$ to get
\be\label{eren1}
\E^{\rm ren} = \E^{\rm f}+\E^{\rm as}
\ee
with the 'finite' part
\be\label{ef} \E^{\rm
  f}=\frac1{2\pi}\sum\limits_{m=-\infty}^{\infty}
\int\limits_{\me}^{\infty}\d
k \ (k^{2}-\me^{2}) \ {\pa\over\pa k} \ \left(\ln f_{m}(ik)-\ln f^{\rm
    as}(ik)\right),  \ee
where it was possible to put $s=0$, and
the 'asymptotic' part
\be\label{eas}
\E^{\rm as}=C_{s}\sum\limits_{m=-\infty}^{\infty}\int\limits_{\me}^{\infty}\d
k \ (k^{2}-\me^{2})^{1-s} \ {\pa\over\pa k} \ \ln f_{m}^{\rm as}(ik)
-\E^{\rm div}. 
\ee
Here, we included the subtraction of $\E^{\rm div}$ according to \Ref{eren}.
The continuation to $s=0$ will yield a finite result because the pole
contributions cancel. This continuation will be done in section
\ref{Sec5}.  But because the expression for $\ln f^{\rm as}$ is quite simple,
this task can be done analytically.

The subdivision \Ref{eren1} of $\E$ is not unique, only the condition
\Ref{diff} has to be satisfied. The inclusion of higher orders into $\ln
f^{\rm as}$ would speed up the convergence of the momentum sum in $\E^{\rm
  f}$, for instance. But we will use $\ln f^{\rm as}$ in the minimal form
obeying \Ref{diff}.

\section{The uniform asymptotic expansion of the Jost function}\label{Sec4}
The uniform asymptotic expansion of the Jost function can be obtained from the
Lippmann-Schwinger equation in much the same way as it was done in the scalar
case \cite{bk96}. We rewrite Eq. \Ref{eq2} in the form 
\[
\left(\begin{array}{rcl}
p_{0}-\me  & & {\pa\over \pa r}-{m\over r}\\
- {\pa\over \pa r}-{m+1\over r}& &p_{0}+\me
\end{array}\right)\Phi(r)={-\delta a(r)\over r}\left(
\begin{array}{rcl}0& &1\\1& &0\end{array}\right)
\Phi(r)\equiv\Delta\P(r)\Phi(r).
\]
The operator on the lhs. can be inverted using the free solutions \Ref{fsol}
and we get the integral equation
\be\label{LS1}
\Phi(r)=\Phi^{0}(r)+\int\limits_{0}^{r}\d r'\,r'  
g(r,r') \ \Delta\P(r')\Phi(r')
\ee
with 
\beq
g(r,r')= -{\pi \over
  2i}\left(\Phi_{J}(r)\Phi^{T}_{H^{(1)}}(r')-
\Phi_{J}(r')\Phi^{T}_{H^{(1)}}(r)\right), \label{green}
\eeq
where $\Phi^{T}$ means the transposed of $\Phi$.
Inserting $\Phi_{J}^{0}$ \Ref{fsol} for $\Phi^{0}(r)$ in the rhs. of Eq.
\Ref{LS1}, this equation determines just the Jost solution. Using the
asymptotic expansion of the Hankel functions and comparing with Eq. \Ref{asj}
we obtain for the Jost function the representation
\be\label{LS2}
f_{m}(k)=1+{\pi\over 2i}\int\limits_{0}^{\infty}\d r\ r\
\Phi^{T}_{H^{(1)}}(r)\Delta\P(r)\Phi(r)\ .
\ee
Equation \Ref{LS1} can be iterated. It turns out that we need all
contributions up to the fourth power in $\Delta\P$  in order to satisfy
condition \Ref{diff}.  Note that in the scalar case the second power had been
sufficient. Iterating Eq. \Ref{LS1} we obtain
\beq
\Phi(r)&=&\Phi^{0}_{J}(r)  \nn\\
&&+\int\limits_{0}^{r}\d r'\, r' g(r,r')\Delta\P(r')\Phi^{0}_{J}(r')
\label{iter}\\
&&+\int\limits_{0}^{r}\d r'\, r'\int\limits_{0}^{r'}\d r'' \,r''
g(r,r')\Delta\P(r')g(r',r'')\Delta\P(r'')\Phi^{0}_{J}(r'') \nn\\
&&+\int\limits_{0}^{r}\d r'\, r' \int\limits_{0}^{r'}\d r'' \,r''
\int\limits_{0}^{r''}\d
r'''\, r''' g(r,r')\Delta\P(r')  \nn\\
&& \hspace{4cm} \times g(r',r'')\Delta\P(r'')g(r'',r''')\Delta\P(r''')
\Phi^{0}_{J}(r''') \nn\\
& &+{\cal O} \left( (\Delta \P )^4 \right). \nn
\eeq
This expression has to be inserted into Eq. \Ref{LS2}. In fact, we need the
logarithm of the Jost function. Therefor the appearing expression must be
expanded again. We write the result as $\ln f_{m}(k)=\sum_{n\ge1}\ln
f^{(n)}_{m}(k)$ where $n$ denotes the power of the operator $\Delta\P$. Up to
the fourth order we obtain (for several details see Appendix A),
\beq 
\ln f^{(1)}_{m}(k)&=&\left({\pi\over 2i}\right)\int\limits_{0}^{\infty}\d r\
r\ \Phi^{T}_{H^{(1)}}(r)\Delta\P(r)\Phi_{J}(r),\label{lne}\\
\ln f^{(2)}_{m}(k)&=&-\left({\pi\over 2i}\right)^{2}
\int\limits_{0}^{\infty}\d r\ r\ \int\limits_{0}^{r}\d r'\ r'\ 
 \Phi^{T}_{H^{(1)}}(r)\Delta\P(r)\Phi_{H^{(1)}}(r)\  \nn\\
&&\hspace{1cm}\times \Phi^{T}_{J     }(r')\Delta\P(r')\Phi_{J    }(r'),
\label{lnz}\\ 
\ln f^{(3)}_{m}(k)&=&2\left({\pi\over 2i}\right)^{3}
\int\limits_{0}^{\infty}\d r\ r\ \int\limits_{0}^{r}\d r'\ r'\ 
\int\limits_{0}^{r'}\d r''\ r''\ 
 \Phi^{T}_{H^{(1)}}(r ) \Delta\P(r  )\Phi_{H^{(1)}}(r)\   \nn \\
&& \hspace{1cm}\times\Phi^{T}_{H^{(1)}}(r' )\Delta\P(r' )\Phi_{J      }(r' )
 \Phi^{T}_{      J}(r'')\Delta\P(r'')\Phi_{J      }(r''),
\label{lnd}\\ 
\ln f^{(4)}_{m}(k)&=&-\left({\pi\over 2i}\right)^{4}
\int\limits_{0}^{\infty}\d r\ r\ \int\limits_{0}^{r}\d r'\ r'\
\int\limits_{0}^{r'}\d r''\ r''\ \int\limits_{0}^{r''}\d r'''\ r'''\  \nn\\ 
&&\hspace{1cm}\times
\Big( 4 
 \Phi^{T}_{H^{(1)}}(r ) \Delta\P(r  )\Phi_{H^{(1)}}(r)\ 
 \Phi^{T}_{H^{(1)}}(r' )\Delta\P(r' )\Phi_{J      }(r' )        \nn    \\
&&\hspace{1.4cm}
 \Phi^{T}_{H^{(1)}}(r'' )\Delta\P(r'' )\Phi_{J      }(r'' )
 \Phi^{T}_{      J}(r''')\Delta\P(r''')\Phi_{J      }(r''') \nn\\ 
&&\hspace{1.4cm}
+2
 \Phi^{T}_{H^{(1)}}(r ) \Delta\P(r  )\Phi_{H^{(1)}}(r)\ 
 \Phi^{T}_{H^{(1)}}(r' )\Delta\P(r' )\Phi_{H^{(1)}}(r' )       \nn       \\
&&\hspace{1.4cm}
 \Phi^{T}_{J      }(r'' )\Delta\P(r'' )\Phi_{J      }(r'' )
 \Phi^{T}_{      J}(r''')\Delta\P(r''')\Phi_{J      }(r''')  \Big),
\label{lnv}
\eeq
where rearrangings of the integration domains had been made. 

Now, because we are interested in the Jost function for imaginary momentum, we
turn from the Bessel functions to the corresponding modified ones. Then we
have to perform the uniform asymptotic expansion of these expressions.  

Before doing this we note that it turned out to be more
convenient not to use the orbital momentum $m$ as the expansion parameter, 
but instead 
\be\label{nu}
\nu=\left\{{m+\frac12 ~~~~~ \mbox{for} ~~~~~  m=0,1,2,\dots  \atop
-m-\frac12 ~~~~~ \mbox{for} ~~~~~ m=-1,-2,\dots  }\right.
\ee
with $\nu=\frac12,\frac32,\dots$ in both cases. 

Then we need uniform asymptotic expansions of the modified Bessel
functions for $\nu\to\infty$, $z$ fixed, of the following type,
\beao
&&\left.{K_{\nu+1/2}(\nu z)\atop I_{\nu+1/2}(\nu z)}\right\}\sim
\sqrt{\pi}^{-\ep}\sqrt{{t\over 2\nu}}\   \left({1-t\over
    1+t}\right)^{\ep/4} \ \e^{\ep\nu\eta(z)}\ \  \exp\Bigg\{ 
     {\frac{-6\,{t^2} - 5\,\epsilon\,{t^3}}{24 \ \nu}}  \\
&&   +{\frac{-4\,\epsilon\,{t^3} - 4\,{t^4} + 5\,\epsilon\,{t^5} + 5\,{t^6}}
     {16 \ \nu^{2}}}   \\
&&  +{\frac{-2160\,{t^4} - 2304\,\epsilon\,{t^5} + 7440\,{t^6} + 
       7695\,\epsilon\,{t^7} - 5400\,{t^8} - 5525\,\epsilon\,{t^9}}{5760 \
       \nu^{3}}}\\
&&+\dots\Bigg\} 
\eeao
with $\ep=\mp 1$ for $K_{\nu}$ respectively $I_{\nu}$, 
$\eta = \sqrt{1+z^2} +\ln (z/(1+\sqrt{1+z^2}))$ and
$t=\left(1+z^{2}\right)^{-{1\over 2}}$ which may be derived from the commonly
known one \cite{abramowitzstegun72} by a corresponding reexpansion. 
Similar expansion for $K_{\nu -1/2}$ and $I_{\nu -1/2}$ are also used.

Now we insert this expansion into the logarithm of the Jost function. Then the
integrations over $r'''$, $r''$ and $r'$ can be carried out successively
by the saddle point method (only equal arguments in the functions $\eta(z)$
yield contributions which do not exponentially decrease for $\nu\to\infty$) as
done in \cite{bk96}. In doing so it becomes apparant that terms up to the
fourth power in $\Delta\P$ contribute to the asymptotic expansion in $\nu$ up
to the order $\nu^{-3}$. The relevant saddle point expansion 
is presented in the Appendix B, see (\ref{saddle}). 
Finally we collect all contributions up to this order
and define
\be\label{lnfas1}
\ln f^{\rm
  as}(ik)=\sum\limits_{n=1}^{3}\sum\limits_{j=n}^{3n}
\int\limits_{0}^{\infty}{\d r\over r}~X_{n,j}{t^{j}\over \nu^{n}}
\ee
with the notation $t=\left(1+(r k/\nu)^{2}\right)^{-{1\over 2}}$.
The coefficients turn out to be
\be\label{Xni}
\begin{array}{rcl}
X_{1,1}&=&\frac{(a\delta)^{2}}{2} \ , \  \ X_{1,3}=
         -\frac{(a\delta)^{2}}{2} \ , \  \nn \\[5pt]
X_{2,2}&=&\frac{1}{4}\delta^2\left( a^{2}-raa'\right) \ , \  
X_{2,4}=\frac{1}{4}\delta^2\left( -3a^{2}+raa'\right) \ , \   
X_{2,6}=\frac{1}{2} (a\delta )^{2}   \nn \\[5pt]
X_{3,3}&=&\frac1 4 \delta^2 
\left(a^{2}-raa'+\frac12r^{2}aa''-\frac 1 2 \delta^2 a^{4}\right)\ , \ 
\nn \\ [5pt]
X_{3,5}&=&  \frac{1}{8}\delta^2 
\left(-\frac{39}{2}a^{2}+7raa'- r^{2} aa''+
6 \delta^2 a^{4}\right) \ , \     \nn \\[5pt]
X_{3,7}&=&\frac{1}{8}\delta^2 \left(35a^{2}-5raa'-5\delta^2 a^{4}\right) \ , \ 
X_{3,9}=\frac{-35}{16}\delta^2 a^{2}.
\nn 
\end{array}\ee
Here, $a$ means the profile function $a(r)$ in \Ref{ma}. Below, when
inserting this expansion into $\E^{\rm as}$, the sum over the orbital momentum
must be performed. There some contributions cancel, for instance those which
are proportional to $\delta$ and $\delta^{3}$. They are not shown in formula
\Ref{Xni}.

\section{The asymptotic part of the \gse}\label{Sec5}
The asymptotic part of the \gse is given by Eq. \Ref{eas} and the expression
\Ref{lnfas1} for the asymptotic expansion of the logarithm of the Jost
function. We rewrite it in the form
\be\label{eas2}
\E^{\rm as}=2 C_{s}
\sum\limits_{\nu=\frac12,\frac32,\dots} \ \int\limits_{\me}^{\infty}\d k \
\left(k^{2}-\me^{2}\right)^{1-s}\ {\pa\over\pa k} 
\int\limits_{0}^{\infty} {\d r\over
      r}\sum\limits_{n=1}^{3}\sum\limits_{j=n}^{3n}X_{n,j}{t^{j}\over \nu^{n}}
\ - \ \E^{\rm div} \ .
\ee
Here, the sum over the orbital momentum $m$ in \Ref{eas} is rewritten as a
sum
over $\nu$, Eq. \Ref{nu}. By means of \Ref{A1} it will be replaced by two 
integrals. The
contribution resulting from the first integral can be calculated explicitely
using formula \Ref{A2}. Together with the explicit expressions for $X_{n,i}$
\Ref{Xni}, after a straightforward calculation, it can be seen to cancel
exactly $\E^{\rm div}$ (for arbitrary profile function $a(r)$).

So we are left with the contribution resulting from the second integral in
\Ref{A1}. There the integration over $k$ can be carried out by means of
formula \Ref{A3}. Then defining
\bea\label{signi}
\Sigma_{n,j}(x)&=&{\Gamma\left(s+\frac{j}{2}-1\right)\over
\Gamma\left(\frac{j}{2}\right)}
\ 
{-i\over x^{j}} \
\int\limits_{0}^{\infty}{\d \nu\over 1+\exp (2\pi\nu)} \\
&& \times 
\left({ (i\nu)^{j-n}\over\left(1+\left(i\nu\over
      x\right)^{2}\right)^{s+\frac{j}{2}-1}}
-
{ (-i\nu)^{j-n}\over\left(1+\left(-i\nu\over
      x\right)^{2}\right)^{s+\frac{j}{2}-1} }
\right)        \nn
\eea
we arrive at 
\bea\label{eas2neu}
\E^{\rm as}=-\frac{1}{\pi}\me^{2}
\sum\limits_{n=1}^{3}\sum\limits_{j=n}^{3n}
\int\limits_{0}^{\infty} {\d r\over r}  X_{n,j} \ \Sigma_{n,j}(r\me).
\eeq

In the functions $\Sigma_{n,j}(x)$ the analytic continuation to $s=0$
has to be performed. For this reason one has to integrate by parts
several times to get rid of the singular denominator. The resulting
expressions are shown in the Appendix C, formulas \Ref{A4}. Now in the
integration over $r$, it is useful to integrate by parts which reduces
to the substitutions $raa'\to-\frac12 a^{2}r\pa_{r}$ and
$r^{2}aa''\to-a'^{2}+\frac12 a^2r\pa_{r}^{2}r$. We note that after
doing this the contribution resulting from $n=2$ vanishes, already
before the integration over $r$ will be carried out. Then the
integrations over $r$ and $\nu$ can be interchanged and after
rescaling $\nu\to \nu r\me$ we get
\be\label{eas3}
\E^{\rm as} =
{-16\over\pi}\int\limits_{0}^{\infty}{\d  r\over r^{3}}\ 
\left\{a(r)^{2} \ g_{1}(r\me)-r^{2}a(r)'^{2} \ g_{2}(r\me)
+a(r)^{4} \ g_{3}(r\me)\right\}\, ,  
\ee
with
\[
g_{i}(x)=  \int\limits_{x}^{\infty} \d \nu \ \sqrt{\nu^{2}-x^{2}} \ f_{i}(\nu)
~~~~ (i=1,2,3)\, .
\]
The functions $f_{i}$ are displayed in the appendix, Eq. \Ref{A5}.  This is
the final formula for $\E^{\rm as}$ for an arbitrary profile function $a(r)$.

For the homogeneous magnetic field inside the flux tube, i.e., for the profile
function \Ref{mah}, the integration over $r$ can be performed
explicitely. After elementary calculations we get
\bea\label{eas4}
\E^{\rm as}&=& {-4\over \pi R^{2}} \Bigg\{
\int\limits_{0}^{R\me} \d \nu \ {\nu^{3}\over 3
  (R\me)^{2}}
\delta^2 \left(f_{1}(\nu)-4f_{2}(\nu)+\frac8{35}\delta^2 f_{3}(\nu)
   \left(\frac{\nu}{m_eR}\right)^4 \right) \\
&&  +  \int\limits_{R\me}^{\infty} \d \nu 
\Bigg[
f_{1}(\nu) \delta^2
\Bigg[
  {\nu^{3}-\sqrt{\nu^{2}-(R\me)^{2}}^{3}\over 3
  (R\me)^{2}}+{\sqrt{\nu^{2}-(R\me)^{2}}\over 2} \nn \\
&&\hspace{3cm} -{(R\me)^{2}\over 2\nu}   \ln
\frac{\left(\nu+\sqrt{\nu^{2}-
(R\me)^{2}}\right)}{m_e R} \Bigg]  \nn \\
&&-4f_{2}(\nu) \delta^2\
  {\nu^{3}-\sqrt{\nu^{2}-(R\me)^{2}}^{3}\over 3
  (R\me)^{2}}                                              \nn \\
&&+f_{3}(\nu)  \delta^4                \nn \\
&&\times\Bigg[
  {8\nu^{7}-\sqrt{\nu^{2}-(R\me)^{2}}\left(8\nu^{6}+4\nu^{4}(R\me)^{2}
+3\nu^{2}(R\me)^{4}-15(R\me)^{6}\right)\over 105
  (R\me)^{6}}                       \nn \\
&&\hspace{2cm}+{\sqrt{\nu^{2}-(R\me)^{2}}\over 2} -{(R\me)^{2}\over 2\nu} 
  \ln\frac{\left(\nu+\sqrt{\nu^{2}-(R\me)^{2}}\right)}
    {m_e R} \Bigg]  \Bigg] \Bigg\}  \, . \nn
\eea
This expression consists of two parts which we write in the form
\be\label{eas5}
\E^{\rm as}=\delta^{2} {e_{1}(R\me)\over R^{2}}+\delta^{4} {e_{2}(R\me)\over
  R^{2}} \, .
\ee
\begin{figure}\unitlength=1cm
\begin{picture}(15,7)
\put(-3,-20){\epsfxsize=20cm\epsffile{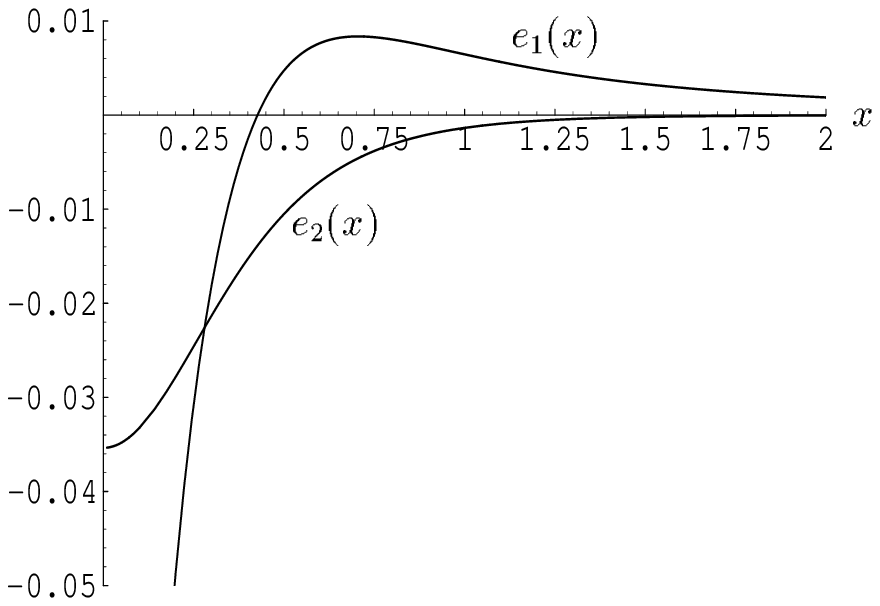}}
\end{picture}
\caption{The functions $e_{1}$ and $e_{2}$ appearing in the asymptotic part of
  the \gse}
\label{fig1}
\end{figure}
Here $e_{1}$ respectively $e_{2}$ 
describe the contributions proportional to the
second respectively 
fourth power of the coupling $\delta$ to the background. They are
shown in Fig. 1. Their behaviour for $x\to 0$ can be calculated from
\Ref{eas4} and we have
\bea\label{easymp}
e_{1}(x)&\sim& {\ln x \over 3\pi}+0.1348 +O(x)\\
e_{2}(x)&\sim&-0.0354 +O(x) \  . \nn
\eea
The logarithmic contribution is just that which was to be expected from the
\hke \Ref{asmod}. 
\section{The 'finite' part of the \gse and numerical results}\label{Sec6}

The finite part of the \gse is defined by Eq. \Ref{ef} together with the
asymptotic expansion of the Jost function, Eq. \Ref{lnfas1}. In general, these
quantities can be calculated only numerically.  We consider here the case of a
homogeneous magnetic field inside the tube as given by Eq.  \Ref{mah}. In that
case the solutions of the field equations are known, they are hypergeometric
functions inside and Bessel functions outside. As these formulas are in
general quite well known resp. easy to derive, we give here only the result.
The notations are close to that in the paper \cite{bv}. For positive orbital
momentum ($m=0,1,\dots$ in formula \Ref{nu}) we have
\bea\label{jfct1} 
f_{\nu}(ik)&=&2\left(\frac{kR}{2}\right)^{\nu+1/2}{\exp
  (-\delta/2)\over\Gamma(\nu+3/2)} 
\left\{ \frac{kR}{2}K_{\nu-\frac12-\delta}(kR) \ {_{1}}F_{1}
\left(1+{(kR)^{2}\over
      4\delta},\nu+\frac32;\delta\right)         \right. \nn\\
&&\left.  \hspace{2cm} + \left(\nu+\frac12\right)K_{\nu+\frac12-\delta}(kR) \
{_{1}}F_{1}\left({(kR)^{2}\over 4\delta},\nu+\frac12;\delta\right)  
\right\} ,    
\eea
and for negative $m$ ($m=-1,-2,\dots$)
\bea\label{jfct2} f_{\nu}(ik)&=&2\left(\frac{kR}{2}\right)^{\nu+1/2}{\exp
  (-\delta/2)\over\Gamma(\nu+3/2)} \nn \\ && \times \left\{
  \frac{kR}{2}K_{\nu-\frac12+\delta}(kR) \ {_{1}}F_{1}\left(\nu+
    \frac12+{(kR)^{2}\over 4\delta},\nu+\frac32;\delta\right)      
   \right. \nn\\
&&\left.  + \left(\nu+\frac12\right)K_{\nu+\frac12+\delta}(kR) \ 
  {_{1}}F_{1}\left(\nu+\frac12+{(kR)^{2}\over
      4\delta},\nu+\frac12;\delta\right) \right\} .  \eea
Thereby the pure Aharonov-Bohm phase is dropped as it does not
contribute to $\E^{\rm f}$. 

The asymptotic part of the Jost function can be obtained explicitely
by carrying out the elementary integrations over $r$ in \Ref{lnfas1}.
Now, having given all ingredients in the integrand of $\E^{\rm f}$,
the remaining task is to perform numerical computations for several
values of the parameters. For this task it turned out to be useful to
integrate by parts and to substitute $k=\sqrt{x}/R$. Then we have
\be\label{ef2}
\E^{\rm
  f}=\frac{-1}{2\pi}\frac{1}{R^{2}}\sum\limits_{\nu=\frac12,\frac32,\dots} \ \
\int\limits_{(R\me)^{2}}^{\infty}\d
x \ \left(\ln f_{\nu}^{+}(ik)+\ln f_{\nu}^{-}(ik)-2\ln f^{\rm
    as}(ik)\right)_{|_{k=\sqrt{x}/R}}. 
\ee
This expression can be calculated numerically. The integration over $x$ is
quite quickly convergent, the sum over $\nu$ not. So, in order to achive a
satisfactory precision for the plots, in depedence of the parameters
$\nu$ must be summed up to   15, for large $x$ even more.  

The general behaviour of $\E^{\rm f}$ as a function of the radius $R$ of the
flux tube is quite smooth. For $R\to 0$ it is proportional to $R^{-2}$. This
can be seen analytically from Eq. \Ref{ef2}. For $R\to\infty$ it is
proportional to $R^{-3}$ which we observed numerically. Having in mind that
the behaviour for $R\to\infty$ is determined by the next \hkk after $a_{2}$ we
conclude from this that $a_{5/2}$ is nonvanishing.  This seems in
contradiction with the general results saying that for manifolds without
boundary half-integer coefficients vanish. But one has to remember, that
higher coefficients contain (at least) squares of derivatives of the
background field which for the presented example leads to undefined
expressions.  Thus, for the higher coefficients the general formulas do not
apply and there is no contradiction at all.

\begin{figure}\unitlength=1cm
\begin{picture}(15,6)
\put(-6,-17){\epsfxsize=25cm\epsfysize=24cm\epsffile{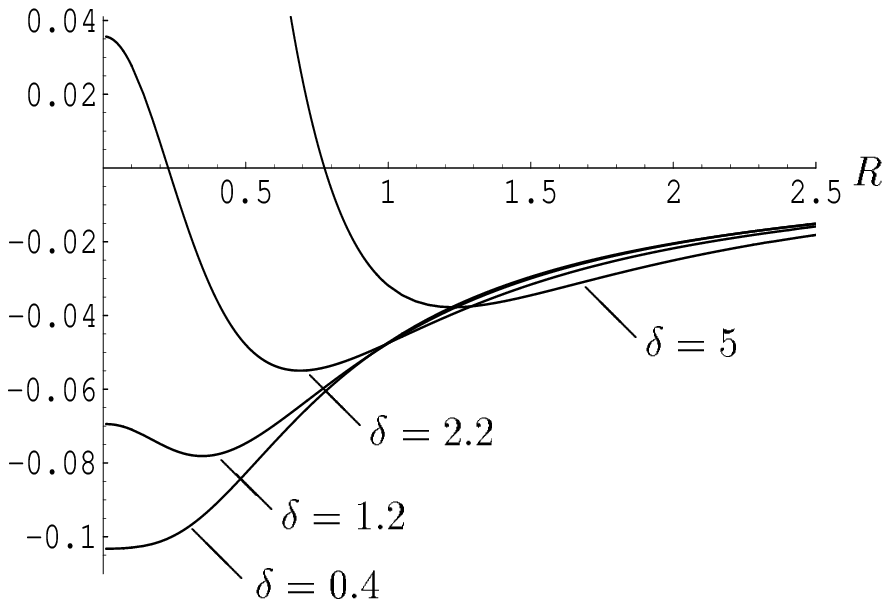}}
\end{picture}
\caption{The function $R^{2}\delta^{-2}\E^{\rm f}(R)$ for several values 
of $\delta$}
\label{fig2}
\end{figure}

In Fig. \ref{fig2} the function $\E^{\rm f}(R)$ is shown multiplied by
$R^{2}\delta^{-2}$ as a function of $R$ for several values of $\delta$.

\begin{figure}\unitlength=1cm
\begin{picture}(15,8)
\put(-6,-18){\epsfxsize=25cm\epsfysize=26cm\epsffile{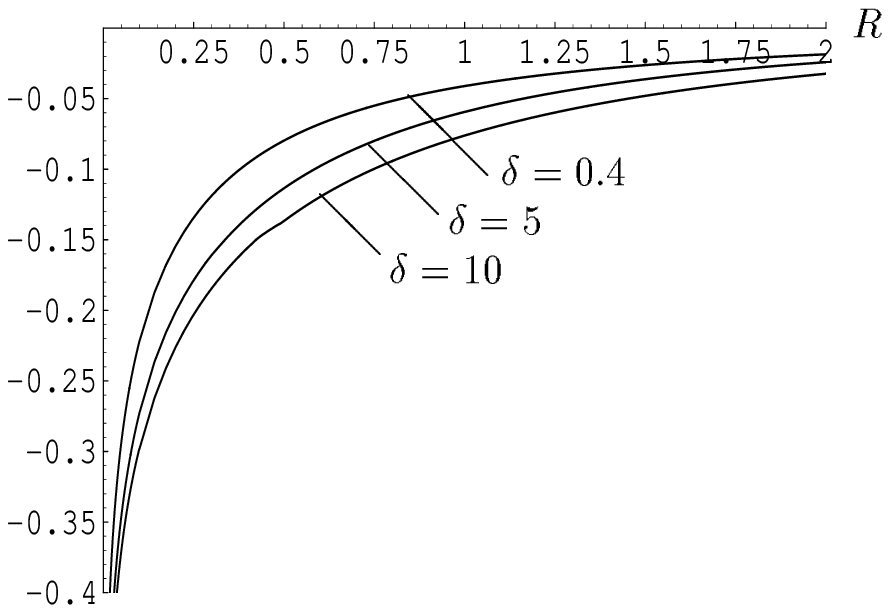}}
\end{picture}
\caption{The  complete \gsep multiplied by $R^{2}\delta^{-2}$ for several
  values of $\delta$}
\label{fig3}
\end{figure}

In Fig. \ref{fig3} the complete \gsek $\E^{\rm ren}(R)$, multiplied by
$R^{2}\delta^{-2}$ is shown for several values of $\delta$. In general, this
function takes only negative values, relatively weakly depending on the flux
$\delta$.  For small $R$, the logarithmic contribution is dominating.

The complete energy is the sum of $\E^{\rm class}$ \Ref{eclass1} and $\E^{\rm
  ren}$ \Ref{eren1}. In Fig.  3, the classical energy would be a straight
horizontal line at $2\pi/\al$. From this it is clear that the complete energy,
remaining a monotone decreasing function of the radius, deviates only slighly
from the classical energy for all values of the radius $R$ except for very
small ones as mentioned in the end of section 2.
 
For large $\delta$, in $\E^{\rm f}$ and $\E^{\rm as}$ the contributions
proportional to $\delta^{4}$ dominate, giving (at last for $0\le R \lsim 1$)
$\E^{\rm f}$ resp. $\E^{\rm as}$ large positive resp. negative values. But
these contributions cancel each other. This was seen in the numerical
calculations. Also, this corresponds to the procedure of adding and
subtracting $\ln f^{\rm as}$ in section 3 which contains terms proportional to
$\delta^{4}$. However, we did not perform a complete investigation of the
behaviour for large $\delta$.

\section{Conclusions}
In this article we have provided a full analysis of the ground state energy 
of the spinor field in the background of a straight magnetic flux tube of 
finite radius $R$. The formalism developed applies in principle to any 
magnetic field with cylindrical symmetry. Assuming that the Jost function 
is known or can be determined numerically, Eqs. (\ref{eas3}) and (\ref{ef2}) 
give the final formulas for this case. We have applied these formulas to the 
case of the magnetic field (\ref{mf}). The final result consists of a 
very explicit "asymptotic" part, Eq. (\ref{eas4}), and a part to be 
determined numerically, Eq. (\ref{ef2}). A detailed numerical analysis 
shows, that the ground state energy turns out be negative, remaining for 
almost all values of the radius $R$ by a factor proportional to the fine 
structure constant $\alpha$ smaller than the classical energy. As a result, 
in the range of applicability of our results, the total energy remains 
positive and, furthermore, does not show a minimum for finite values of $R$ 
given a fixed flux. The magnetic string thus remains unstable also when 
including quantum corrections into the total energy.

An interesting question is on the general dependence of the \gse on the
specific background chosen here.  First we remark, that we are in agreement
with \cite{dunne} with respect to the \gse beeing a small addendum to the
classical energy (as fas as $R$ is not too small). Secondly, from the explicit
formulas like \Ref{eas3} for $\E^{\rm as}$ we expect that the dependence on
the shape of the magnetic field inside the flux tube will be weak. This will
be so at last for sufficiently smooth background fields. When, however, the
background becomes singular, this may change.  As an example one can consider
the \gse of a massive spinor field with bag boundary conditions on a sphere
calculated in \cite{ebk98} showing as function of the radius even changes of
the sign.  Another consequence of our calculations is that it seems impossible
to shrink the radius of the string to zero bcause of the logarithmic
singularity \Ref{asmod} appearing in that case. In view of this, it would be
interesting to reconsider earlier investigations in the background of the
infinitely thin string whereby we admit that it might well happen that this
singularity can be absorbed into some counterterm.

Even if we have found a negative answer within the class of examples
(\ref{mf}), the results presented can be a starting point to consider further
the question if inhomogeneous magnetic fields can minimize the energy for
fixed flux. Furthermore, it seems possible to include other aspects as for
example external electric fields and the anomalous magnetic moment.  Also, the
techniques developed here, are suited for the calculation of the fermionic
contribution to the vacuum polarisation in the background of the
Nielsen-Olesen vortex, the Z-string or in a chromomagnetic background.
 
It might be interesting to compare the \gse found here, which is complely
negative, with the effective potential in a homogeneous background field which
was mentioned in the introduction. For small fields it takes positive values
which are important for the electroweak symmetry restoration. So our result
suggests that this might be altered by a finite spatial extension of the
field.

\section*{Acknowledgments}
KK has been supported by the DFG under contract number Bo 1112/4-2.

\section*{Appendix A: Perturbation theory for the 
logarithm of the Jost function}
In this appendix we will derive the expansion for the 
logarithm of the Jost function, Eq. (\ref{LS2}). The first step is to use
Eq. (\ref{iter}) in Eq. (\ref{LS2}). The Jost function itself up to the 
fourth power in the perturbation $(\Delta ({\cal P})$ reads
\beq
f_m (k) =: 1 +x_1 +x_2+x_3+x_4+{\cal O} \left( (\Delta {\cal P}) ^5 \right),
\label{B1}
\eeq
with the definitions
\beq
x_1&=& \pre \intre \fhe \dpe \fje , \label{B2}\\
x_2 &=& \pre \intre \intrz \fhe \dpe  g (r_1,r_2) \dpz \fjz ,\label{B3}\\
x_3 &=& \pre \intre \intrz \intrd  \fhe \dpe 
               \nn\\
& & \hspace{4cm} g(r_1,r_2) \dpz 
           g(r_2,r_3) \dpd \fjd , \label{B4}\\
x_4 &=& \pre \intre \intrz \intrd \intrv \fhe \dpe \nn\\
& & \hspace{2cm} g(r_1,r_2) 
          \dpz g(r_2,r_3) \dpd g(r_3,r_4) \dpv \fjv . \label{B5}
\eeq
We will need the combinations 
\beq
\ln f_m (k) = \ln f_m ^{(1)} (k) +\ln f_m ^{(2)} (k)+\ln f_m ^{(3)} (k) +
\ln f_m ^{(4)} (k) +{\cal O} \left( (\Delta ( {\cal P})^5 \right) , 
\label{B6}
\eeq
with 
\beq
\ln f_m^{(1)} &=& x_1 ,\label{B7}\\
\ln f_m^{(2)} &=& x_2 -\frac 1 2 x_1^2 , \label{B8}\\
\ln f_m^{(3)} &=& \frac 1 3 x_1^3 -x_1 x_2 +x_3 , \label{B9}\\
\ln f_m^{(4)} &=& -\frac 1 4 x_1^4 +x_1^2x_2 -\frac 1 2 x_2^2-x_1 x_3 +x_4 .
\label{B10} 
\eeq
Let us consider $\ln f_m (k)$ order by order. The first order 
$\ln f_m ^{(1)} (k)$ is already given by its definition, Eqs. (\ref{B7}) 
and (\ref{B2}). For the calculation of $\ln f_m ^{(2)} (k)$ some 
manipulations are needed. The main trick, also for the calculation 
of the higher orders, is the rearrangement of integration domains.
At the beginning we will give details, later on only an idea of the 
single steps is given.

Using Eq. (\ref{green}) one obtains
\beq
\ln f_m ^{(2)} (k) &=& \pre^2 \left\{\intre 
\intrz \fhe \dpe \fje \fhz \dpz \fjz 
\right.\nn\\
& &-\intre \intrz \fhe \dpe \fhet \fjzt \dpz \fjz \nn\\
& &\left.-\frac  1 2 \left[ \intre \fhe \dpe \fje \right]^2 \right\}\nn
\eeq
The first and third term combine to give 
\beq
& &\frac 1 2 \pre ^2 \intre \fhe \dpe \fje \times \nn\\
& &\left\{\intrz \fhz \dpz \fjz - \int_{r_1}^{\infty} dr_2 \, 
\fhz \dpz \fjz \right\} .\nn
\eeq
Next the integration domains may be rearranged,
\beq
\intre \int_{r_1}^\infty dr_2  = \int_0^{\infty} dr_2 \int_0^{r_2} dr_1 \nn
\eeq
and changing finally the name of the variable, $r_1 \leftrightarrow r_2$, 
one arrives at Eq. (\ref{lnz}). 

When calculating the higher orders it is extremely helpful to systematically
use the lower orders already obtained. So for the next order we start with
\beq \ln f_m ^{(3)} (k) = x_3 -\frac 1 6 (\ln f_m ^{(1)} ) ^3 - (\ln f_m
^{(1)} )(\ln f_m ^{(2)} ) , \nn \eeq where \beq x_3 = x_{3,1} +x_{3,2}
+x_{3,3} +x_{3,4} \nn \eeq consists of the pieces \beq
x_{3,1} &=& \pre ^3 \intre \intrz \intrd \fhe \dpe \fje \nn\\
& &\hspace{2cm}\fhz \dpz \fjz
\fhd \dpd \fjd \nn\\
x_{3,2} &=& -\pre ^3 \intre \intrz \intrd \fhe \dpe \fje \nn\\
& &\hspace{2cm}\fhz \dpz \fhzt
\fjdt \dpd \fjd \nn\\
x_{3,3} &=& -\pre ^3 \intre \intrz \intrd \fhe \dpe \fhet \nn\\
& &\hspace{2cm}\fjzt \dpz \fjz
\fhd \dpd \fjd \nn\\
x_{3,4} &=& \pre ^3 \intre \intrz \intrd \fhe \dpe \fhet\nn\\
& &\hspace{2cm} \fjzt \dpz \fhzt \fjdt \dpd \fjd \, . \nn \eeq 
Cancellations occur
due to the identity \beq \intre \intrz ... \int_0^{r_n}dr_{n+1} f(r_1) ...
f(r_{n+1}) = \frac 1 {(n+1)!} \left[ \int_0^\infty dr f(r) \right] ^{n+1} ,
\label{indu} \eeq which can be proven by induction. It shows \beq x_{3,1}
-\frac 1 6 (\ln f_m^{(1)}) ^3 =0 .\nn \eeq To manipulate the contribution
$(\ln f_m^{(1)}) (\ln f_m ^{(2)})$ integrals are spitted according to \beq
\int_0^\infty dr_3 = \int_0^{r_2} dr_3 + \int_{r_2}^\infty dr_3 \label{split}
\eeq and identities of the kind \beq \int_0^r dr_1 \, \int_{r_1}^\infty dr_2 =
\int_0^r dr_2 \int_0^{r_2}dr_1 + \int_r^\infty dr_2 \int_0^r dr_1
\label{retwo} \eeq are used. One arrives at \beq -(\ln f_m^{(1)}) (\ln
f_m^{(2)}) = -x_{3,2} -x_{3,3} +x_{3,4} \nn \eeq ending up with Eq.
(\ref{lnv}).

Finally, the last order we will need can be written as 
\beq
\ln f_m^{(4)} (k) &=& x_4 -\frac 1 2 (\ln f_m^{(2)} (k) )^2 
-\frac 1 2 (\ln f_m^{(1)} (k) ^2 ( \ln f_m ^{(2)} (k) ) \nn\\ 
& &-(\ln f_m^{(1)} (k) ) ( \ln f_m ^{(3)} (k) ) -
\frac 1 {24} (\ln f_m^{(1)} ) ^4 .\nn
\eeq
The contribution 
\beq
x_4 = \sum_{i=1}^8 x_{4,i} \nn
\eeq
consists of 
\beq
x_{4,1}&=& \intre \intrz \intrd \intrv \fhe \dpe \fje \nn\\
& &\fhz \dpz \fjz \fhd \dpd \fjd \fhv \dpv \fjv \nn\\
x_{4,2}&=&- \intre \intrz \intrd \intrv \fhe \dpe \fje \nn\\
& &\fhz \dpz \fjz \fhd \dpd \fhdt \fjvt \dpv \fjv \nn\\
x_{4,3}&=& - \intre \intrz \intrd \intrv \fhe \dpe \fje \nn\\
& &\fhz \dpz \fhzt \fjdt \dpd \fjd \fhv \dpv \fjv \nn\\
x_{4,4}&=& \intre \intrz \intrd \intrv \fhe \dpe \fje \nn\\
& &\fhz \dpz \fhzt \fjdt \dpd \fhdt \fjvt \dpv \fjv \nn\\
x_{4,5}&=&-\intre \intrz \intrd \intrv \fhe \dpe \fhet \nn\\
& &\fjzt \dpz \fjz \fhd \dpd \fjd \fhv \dpv \fjv \nn\\
x_{4,6}&=& \intre \intrz \intrd \intrv \fhe \dpe \fhet \nn\\
& &\fjzt \dpz \fjz \fhd \dpd \fhdt \fjvt \dpv \fjv \nn\\
x_{4,7}&=& \intre \intrz \intrd \intrv \fhe \dpe \fhet \nn\\
& &\fjzt \dpz \fhzt \fjdt \dpd \fjd \fhv \dpv \fjv \nn\\
x_{4,8}&=&-\intre \intrz \intrd \intrv \fhe \dpe \fhet \nn\\
& &\fjzt \dpz \fhzt \fjdt \dpd \fhdt \fjvt \dpv \fjv \, . \nn  
\eeq
Eq. (\ref{indu}) shows 
\beq
x_{4,1} -\frac 1 {24} (\ln f_m^{(1)} (k) ) ^4 =0.\nn
\eeq
With the help of rearrangements as (\ref{split}), (\ref{retwo}), and 
similar ones, it can be shown that 
\beq
-(\ln f_m ^{(1)} (k) ) (\ln f_m^{(3)} (k) ) &=& 
-2x_{4,7} +4 x_{4,8} -2 x_{4,4} ,\nn\\
-\frac 1 2 (\ln f_m^{(1)} (k) ) ^2 (\ln f_m^{(2)} (k) ) &=& 
-x_{4,5} +x_{4,7} -x_{4,3} -x_{4,8} +x_{4,4} -2x_{4,2} ,\nn
\eeq
and
\beq
\lefteqn{
-\frac 1 2 (\ln f_m^{(2)} (k) )^2 = -x_{4,6} }\nn\\ 
& &-2 \intre 
\intrz \intrd \intrv \fhe \dpe \fhet \nn\\
& &\fhz \dpz \fhzt \fjdt \dpd \fjd \fjvt \dpv \fjv .\nn
\eeq
Putting all pieces together one arrives at (\ref{lnv}). 

\section*{Appendix B: Saddle point expansion of integrals}
For the derivation of Eqs. (\ref{lnfas1}) and (\ref{Xni}) repeated use
of saddle point expansions was made. The relevant result is stated in
this Appendix. 

For $\nu\to \infty$ one obtains the following asymptotic expansion,
\beq
\int_0^r dr' \phi (r') e^{\nu \varphi (r')} =
e^{\nu \varphi (r)} \sum_{k=1}^{\infty} h_{k-1} \nu^{-k} , \label{saddle}
\eeq
where the needed leading terms of the expansion are
\beq
h_0 &=& \frac{\phi (r)}{\varphi ' (r)} ,\nn\\
h_1 &=& \frac{\phi (r) \varphi '' (r)}{(\varphi ' (r))^3} 
             -\frac{\phi' (r)}{(\varphi ' (r))^2} ,\nn\\
h_2 &=& \frac{\phi '' (r)}{(\varphi ' (r))^3} -
          \frac{3\phi' (r) \varphi '' (r)}{(\varphi ' (r) )^4} +
        \frac{3 \phi (r) (\varphi '' (r))^2}{(\varphi ' (r) ) ^5 } -
         \frac{\phi (r) \varphi ''' (r)}{(\varphi ' (r) ) ^4 }. \nn
\eeq

\section*{Appendix C: Representation of sums as integrals} 
Here we display some formulas used in this paper, where sums are 
replaced by integrals. 

The sum over half integer numbers can be represented by two integrals, the
necessary analytical properties of the function $f(\nu)$ beeing assumed:
\be\label{A1}
\sum\limits_{l=0}^{\infty}f(l+\frac{1}{2})=\int\limits_{0}^{\infty}\d
\nu~f(\nu)+\int\limits_{0}^{\infty}{\d \nu\over
  1+\e^{2\pi\nu}}{f(i\nu)-f(-i\nu)\over i}.
\ee
The first part of $\E^{\rm as}$ in section \ref{Sec5} can be calculated using
\be\label{A2} \int_{0}^{\infty}d\nu\int_{m}^{\infty}dk(k^{2}-m^{2})^{1-s}
{\partial\over\partial k}{t^{i}\over\nu^{n}}=-{m^{2-2sn}\over 2}
{\Gamma(2-s)\Gamma(\frac{1+i-n}{2})\Gamma(s+\frac{n-3}{2})\over (rm)^{n-1} \ 
  \Gamma(i/2)}.  \ee
The integration over $k$ in formula \Ref{eas2} can be done using
\be\label{A3}
\int_m^\infty dk(k^2-m^2)^{1-s}{\partial\over\partial
  k}t^{i}=-m^{2-2s}{\Gamma(2-s)\Gamma(s+\frac{i}{2}-1)\over
\Gamma(\frac{i}{2})}{\left({\nu\over mr}\right)^{i}\over 
\left(1+\left({\nu\over mr}\right)^{2}\right)^{s+\frac{i}{2}-1}}.
\ee
The functions $\Sigma_{n,j}(x)$, \Ref{signi} can be written as
\be\label{A4}
\Sigma_{n,j}(x)= {4\over x^{2}} \ \int\limits_{x}^{\infty}\d \nu \
\sqrt{\nu^{2}-x^{2}} \ f_{n,j}(\nu), 
\ee
for $n=1$, $j=1,3$ and $n=3$, $j=3,5,7,9$ with
\beao
f_{1,1}(\nu)&=&-{1\over 1+\exp  (2\pi\nu)}  \nn \\
f_{1,3}(\nu)&=&-\left({\nu\over 1+\exp  (2\pi\nu)}\right)'  \nn \\
f_{3,3}(\nu)&=&\left(\frac{1}{\nu} \ {1\over 1+\exp (2\pi\nu)}\right)' \nn \\
f_{3,5}(\nu)&=&\frac13\left(\frac{1}{\nu}\left({\nu\over 1+
\exp (2\pi\nu)}\right)'\right)'  \\
f_{3,7}(\nu)&=&\frac1{15}\left(\frac{1}{\nu}\left(\frac{1}{\nu}
    \left({\nu^3\over
        1+\exp  (2\pi\nu)}\right)'\right)'\right)'  \nn \\
f_{3,9}(\nu)&=&\frac1{105}\left(\frac{1}{\nu}\left(\frac{1}{\nu}
\left(\frac{1}{\nu}\left({\nu^5\over
        1+\exp  (2\pi\nu)}\right)'\right)'\right)'\right)'  \ .  \nn 
\eeao
For $n=2$ the formulas are slightly more explicit. They read
\beq
\Sigma_{2,2} &=& -\frac 1 {x^2} \ln \left( 1+e^{-2\pi x}\right),\nn\\
\Sigma_{2,4} &=& \frac{\pi} x \frac 1 {1+e^{2\pi x} } , \nn\\
\Sigma_{2,6} &=& \frac{3\pi}{4x} \frac 1 {1+e^{2\pi x} } -
\frac{\pi^2} 2 \frac{e^{2\pi x}} {(1+e^{2\pi x})^2} .\nn
\eeq
The functions $f_{i,j}$ build the ingredients for the function $g_i (x)$ 
in Eq. (\ref{eas3}). Explicitely we find
\bea\label{A5}
f_{1}(x)&=&\frac12 f_{1,1}(x)-\frac12 f_{1,3}(x)+\frac1{4}
f_{3,3}(x)-\frac{39}{16} f_{3,5}(x)+\frac{35}{8} f_{3,7}(x)-\frac{35}{16}
f_{3,9}(x) \nn\\
&& -\frac12
x\pa_{x}\left(-\frac1{4}f_{3,3}(x)+\frac7{8}f_{3,5}(x)-
\frac5{8}f_{3,7}(x)\right)\nn\\
&&+\frac12
x\pa_{x}^{2}\left(\frac{x}{8}f_{3,3}(x)-\frac{x}{8}f_{3,5}(x)\right), \nn\\
f_{2}(x)&=&\frac1{8}\left(f_{3,3}(x)-f_{3,5}(x)\right), \\
f_{3}(x)&=&-\frac1{8}\left(f_{3,3}(x)-6f_{3,5}(x)+5f_{3,7}(x)\right) . \nn
\eea

\end{document}